\documentclass[12pt]{article}
\usepackage{amssymb}
\usepackage{epsfig}

\catcode`\@=11
\def\numberbysection{\@addtoreset{equation}{section}
        \def\theequation{\thesection.\arabic{equation}}}
\def\beq{\begin{equation}}
\def\eeq{\end{equation}}
{\numberbysection
\begin{document}
\begin{titlepage}
\begin{center}
\hfill DFF 1/11/97 \\
\vskip 1.in
{\Large \bf Exact Results in Chern-Simons Supergravity}
\vskip 0.5in
P. Valtancoli
\\[.2in]
{\em Dipartimento di Fisica dell' Universita', Firenze \\
and INFN, Sezione di Firenze (Italy)}
\end{center}
\vskip .5in
\begin{abstract}
We propose an extension of a recent non-perturbative method suited for
solving the $N$-body problem in $(2+1)$-gravity to the case of
Chern-Simons supergravity. Coupling with supersymmetric point
particles is obtained implicitly by extending the $DJH$ matching
conditions of gravity.

The consistent solution of the interacting case is obtained by
building a general non-trivial mapping, extending the superanalytic
mapping, between a flat polydromic $X^M$ supercoordinate system and a  
physical one $x^N$, representing the $DJH$ matching conditions around
the superparticles. We show how to construct such a mapping in terms
of analytic functions, and we give their exact expressions for the 
two body case. The extension to the $N$ body case is also discussed. 

In the Minkoskian coordinates the superparticles move freely, and in
particular the fermionic coordinates $\Theta(\xi^N_{(i)})$ are
constants, whose values can be fixed by using the monodromy properties.
While the bosonic part of the supergeodesic equations are obtained, as
in gravity, by measuring the bosonic distance in Minkowskian 
space-time, we find that the fermionic geodesic equations can be
defined only by requiring that a non-perturbative divergence of the 
$X^M = X^M( x^N)$ mapping cancels out on the world-lines of the superparticles.
\end{abstract}
\medskip
\end{titlepage}
\pagenumbering{arabic}

\section{Introduction}

Recently, a series of paper have renewed a considerable interest for
the study of the classical problem of $N$ particles interacting with 
$(2+1)$-gravity  
\cite{star}-\cite{deser}-\cite{quan}-\cite{gott}-\cite{capp}-\cite{bell}-
\cite{ciaf}-\cite{well}. In this paper we have investigated the
supergravity case, following also  ref. \cite{cvaz}, applying an
improved version of the method which enabled us to
solve the classical $N$-body problem in (2+1)-gravity \cite{ciaf}. 
The motivation for such study is twofold, since firstly it gives an
opportunity to study non-perturbatively for the first time a classical
interacting solution in a supergravity theory, and secondly since it
permits to study the spin as an anticommuting variable, already 
built in in the superalgebra. In fact this interest came to us 
after the study of spinning particles \cite{clem}-\cite{spin}, 
which are affected by the $CTC$ problem \cite{sctc}.
In that case we found that closed temporal curves appear inside a $CTC$
horizon, surrounding the particle site, whose radius is of the order of $S^2$,
the square of the spin. So what happens when instead of treating the
spin as a classical variable we introduce it as a
pseudoclassical variable ? The natural framework in which this idea
could be tested is $(2+1)$ supergravity ( see \cite{gate} for a review ), 
where we are forced to write equations in the framework of
pseudoclassical dynamics \cite{bere}. 

The particular problem of $N$ superparticles coupled to supergravity 
\cite{cvaz} seems particular interesting because captures many of the 
simplifying features of $(2+1)$ gravity. Let us briefly recall them: 
the space-time of a point source is a cone with deficit angle and can
be represented as a Minkoskian space-time with an excised region 
\cite{deser}. Moreover the problem of finding the metric of $N$ bodies
can be reduced to find a
mapping between a multivalued Minkowskian coordinate system $X^a$ and
a physical coordinate system $X^a = X^a(x^b)$ \cite{capp}. 
In a particular instantaneous gauge it was found
in ref. \cite{ciaf} that such a Minkowskian coordinate system can be
represented as a sum of polydromic analytic and antianalytic functions, with
branch points at the particle sites . Then integrating the geodesic equations
was found to be equivalent to measure the bosonic distance $X^a(\xi_i)
- X^a(\xi_j)$ between the particles in the Minkowskian coordinates.

Since the same simplifications that permit to solve $(2+1)$-gravity
still hold in the case of $(2+1)$-supergravity, we were encouraged to 
understand the quite involved equations of supergravity 
\cite{gate}-\cite{bagg} 
as a non-trivial, polydromic mapping between
supercoordinates $X^M = X^M( x^N )$, where now the index $N$ covers
also the anticommuting variables \cite{suma}-\cite{roge}. As a first
step we find that the simplest way to couple a superparticle to
supergravity is to introduce a composite monodromy matrix, carrying
not only Lorentz cuts but also supersymmetric cuts. In particular, in 
addition to the bosonic constants of motion, which appear as
parameters into the definition of the monodromy matrix, we introduce 
fermionic constant parameters, induced by the supersymmetric source,
which are useful to parameterize the solution for the gravitino
field. These constants describe also the free motion of the
superparticle in the flat superspace $X^M$, and in particular the
fermionic coordinate $\Theta(\xi^N)$ of the superparticle is constant, whose  
value is determined by the monodromy matrix.

Then we are able to show that the polydromic supercoordinate
transformation $X^M = X^M (x^N)$ can be represented as a sum of
analytic and anti-analytic functions, as in the case of gravity
\cite{ciaf}, whose form resembles the superanalytic
mapping. The bosonic part of this mapping is simply
dictated by the solution of the corresponding problem in
$(2+1)$-gravity. The non-triviality of supergravity is hidden in
finding a solution to a polydromic mapping for the gravitino field,
which however can be solved exactly for the two body problem.

To explain our results on the superspace mapping it is better to review
the results of gravity, which represents the pure bosonic part
of it, if we avoid all the dependence from the anticommuting variables. 
In the case of massive spinless particles coupled to gravity 
the polydromy is well defined at the particle site, since it can be
reduced to a pure rotation. In the case of spinning particles, the 
definition of such a mapping is more problematic, due to the
translation monodromy giving rise, for a static particle, to a 
logarithmic cut in the $T$ variable. Therefore the $X^a$ variables are
no more single valued at the particles sites, as noted in \cite{capp}.

Moreover, for a static spinning source it was surprisingly found
another non-perturbative divergence of the $Z$-mapping 
\cite{well}-\cite{spin}, apparently without any motivation although in
our instantaneous gauge it is related to the $CTC$ problem
\cite{sctc}. In the framework of supergravity, we find that the 
$X^M=X^M(x^N)$ mapping is plagued with the same divergences of the
spinning case, at least with our gauge choice. It appears a
logarithmic divergence due to the spin, and again
the same type of divergence analogous to the $Z$-mapping divergence of
the spinning case. However, in this case the superspace geometry helps
in giving an interesting interpretation to it. According to our
computations, we find that the $\theta$ terms carry an analogous 
divergence, and that it is possible to cancel the fermionic residue of
such divergence. Requiring such cancellation at the superparticle site
is then equivalent to integrate the fermionic part of the
supergeodesics equations in the present formalism, as measuring the
bosonic distance between two particles in Minkowskian coordinates was 
equivalent to integrate the geodesic equations in gravity. 

Therefore to study superparticle dynamics it is not possible to truncate 
the supermapping to the $\theta = 0$ part. We also give a general
proof that the mechanism of cancellation of divergences is
self-consistent at all orders, and we determine the general solution
for the fermionic geodesic equation. Finally we are also able to
integrate the bosonic trajectories of the superparticles and give
the classical scattering angle.
  
The contents of the paper are as follows. In the introductory Section 2 we
recall the Chern-Simons formalism for introducing $(2+1)$-supergravity
and, following \cite{ciaf}, we define the mapping problem from a multivalued
flat supercoordinate system to single valued coordinates 
in a instantaneous gauge. In Sec. 3 we treat the 
two superbody problem, and in particular, we give the non perturbative
interacting solution for the $X^M = X^M(x^N)$ mapping and the motion of the 
superparticles. We also outline how to solve the many body case. 
Finally we discuss our results in  Section 4, giving a few concluding remarks. 

\section{Supergravity Lagrangian}

The $N=1$ Super-Poincar\`e algebra $ISO(2,1|2)$ is given in terms of
real generators as follows

\begin{eqnarray}
& [ J^a , J^b ] & =  \epsilon^{abc} J_c \ \ \ \ [ J^a , Q_{\alpha} ] = -
\frac{1}{2} \gamma^a_{,\alpha}{}^\beta Q_{\beta} \nonumber \\
& [ J^a , P^b ] & =  \epsilon^{abc} P_c \ \ \ \ \{ Q_{\alpha} , Q_{\beta} \} = 
- \frac{1}{4} {( \gamma^a C^{-1} )}_{\alpha \beta} P_a \nonumber \\
& [ P^a , P^b ] & =  [ P^a , Q_{\alpha} ] = 0, 
\label{a1} \end{eqnarray}

where $C^{\alpha \beta}$ is the charge conjugation matrix $C^{12} = -
C^{21} = 1$ and $J^a = \frac{1}{2} \epsilon^{abc} J_{bc}$.

Following Witten \cite{witt}, we recall how to formulate $(2+1)$ gravity
theories in terms of a topological Chern-Simons action. In general it
can be written for a compact gauge group $G$ as
\beq I_{CS} = \frac{1}{2} \ \int_M \ Tr \left( A \wedge d A + \frac{2}{3} A
\wedge A \wedge A \right) \ , \label{a2} \eeq
with $A = A^a T_a$ a Lie-algebra-valued one form. The kinetic term is
dependent on the choice of the trace operation. For compact Lie groups
this is usually chosen as $d_{ab} = Tr (T_a T_b )$.  
More generally, $d_{ab}$ is an ad-invariant quadratic form on the Lie 
algebra. To write a consistent action we must require that this
quadratic form is non-degenerate. For the general case of the group 
$ISO(d-1,1)$, which is non semi-simple, there is no candidate for $d_{ab}$
because the natural quadratic Casimir $C_1 = P^a P_a$ is
degenerate. In the special $d=3$ case however there is another
quadratic Casimir $C_2 = J^a P_a$ which induces a natural
non-degenerate quadratic form 
\beq  < J_a , P_b > \ = \ \delta_{ab} , \ \ \ \ \ 
< J_a , J_b > \ = \ < P_a , P_b >  = 0 .\label{a3} \eeq
Then the $ISO(2,1)$ action, endowed with the bilinear form (\ref{a3}),
corresponds to the action of $(2+1)$-gravity. 

What happens in $(2+1)$ supergravity ? The supersymmetric generators
do not contribute to the mass Casimir $C_1 = P_a P^a$, 
while the second Casimir is modified as 
\beq C_2 = P_a J^a + \overline{Q}^\alpha Q_\alpha , \label{a4} \eeq
where $ \overline{Q}^\alpha = Q_\beta C^{\beta\alpha} $. It induces a
corresponding quadratic form $d^{AB}$, with which it is possible to 
write down the supergravity Lagrangian as a Chern-Simons theory on
the $N=1$ Super-Poincar\`e algebra $ISO(2,1|2)$. Defining
\beq A_\mu = e^a_\mu P_a + \omega^a_\mu J_a +
\overline{\psi}^\alpha_\mu Q_\alpha \label{a5}, \eeq
the Chern-Simons action for $N=1$ supergravity is defined by (\ref{a2}).
From $d_{AB}$ defined from (\ref{a4}), we obtain the usual
action for simple supergravity

\begin{eqnarray}
 {\cal L} & = &  - \frac{1}{2} \int_M \ d^3 x \ \epsilon^{\mu\nu\rho}
[ e^a_\mu ( \partial_\nu \omega_{a\rho} - \partial_\rho \omega_{a\nu}
+ \epsilon_{abc} \omega^b_\nu \omega^c_\rho ) + \nonumber \\
& + & \frac{1}{2} \overline{\psi}_\mu ( \partial_\nu + \frac{1}{2}
\omega^a_\nu \gamma_a ) \psi_\rho ]. \label{a6} \end{eqnarray}

The equations of motion of $(2+1)$-supergravity reveal the presence of a
distributed torsion source, due to the presence of the gravitino
field

\begin{eqnarray}
\partial_\mu \omega_{a\nu} & - & \partial_\nu \omega_{a\mu} +
\epsilon_{abc} \omega_\mu^b \omega_\nu^c = 0 \nonumber \\
\partial_\mu e_{a\nu} & - & \partial_\nu e_{a\mu} + 2 \epsilon_{abc}
e^b_\mu \omega^c_\nu + \frac{1}{8} \overline{\psi}_\mu \gamma_a
\psi_\nu - \frac{1}{8} \overline{\psi}_\nu \gamma_a \psi_\mu = 0
\nonumber \\
( \partial_\mu & + & \frac{1}{2} \omega_\mu^a \gamma_a ) \psi_\nu - 
( \partial_\nu + \frac{1}{2} \omega_\nu^a \gamma_a ) \psi_\mu = 0 .
\label{a7} \end{eqnarray}

These can be simplified by going to a singular gauge, similar to what has
been discussed in \cite{capp} 
for gravity, and choosing a polydromic dreibein and  
gravitino fields such that $\omega_{a\mu} = 0 $ locally, from which
one obtains the simplified equations:

\begin{eqnarray}
\omega^a_\mu & = & 0 \nonumber \\
\partial_\mu e^a_\nu & - & \partial_\nu e^a_\mu + \frac{1}{8}
\overline{\psi}_\mu \gamma^a \psi_\nu - \frac{1}{8}
\overline{\psi}_\nu \gamma^a \psi_\mu = 0 \nonumber \\
\partial_\mu \psi_\nu & - & \partial_\nu \psi_\mu = 0. 
\label{a8} \end{eqnarray}

The equation for $\psi_\mu$ implies that the gravitino field is a pure
gauge, being the derivative of a two component Majorana spinor $\psi$:
\beq \psi_\mu = \partial_\mu \psi. \label{a9} \eeq
Analogously, the equation for the dreibein can be simplified, after an
integration by parts of the gravitino contribution, as
\beq \partial_\mu ( e^a_\nu + \frac{1}{16} \overline{\psi} \gamma^a
\stackrel{\leftrightarrow}{\partial_\nu} \psi ) - 
\partial_\nu ( e^a_\mu + \frac{1}{16}\overline{\psi} \gamma^a 
\stackrel{\leftrightarrow}{\partial_\mu} \psi ) = 0. \label{a10} \eeq
Therefore we can define a Minkowskian bosonic coordinate $X^a$
\beq e^a_\mu + \frac{1}{16} \overline{\psi} \gamma^a 
\stackrel{\leftrightarrow}{\partial_\mu} \psi
= \partial_\mu X^a, \label{a11} \eeq
in analogy to gravity.

We have been able to reconstruct, starting from the equations of
motion, two primitive functions, one bosonic $X^a$ and the other
fermionic $\psi$. This corresponds to the well known property that $(2+1)$
supergravity is a pure topological theory, i.e. there are no dynamical
degrees of freedom, no free gravitons or gravitinos. The only physical
degrees of freedom are those of the external sources ( the simplest ones
are a set of $N$ point superparticles ) or topological degrees of freedom. In
this paper we will investigate the coupling of $N$-point sources to 
supergravity, since in this case the supercurvature is concentrated on
the point sources. To allow the coupling of a superparticle to
supergravity we must require certain polydromic conditions on the 
$(X^a,\psi)$-mapping, similar to the Deser, Jackiw and 't Hooft ($DJH$)
matching conditions \cite{deser}. As a consequence of eqs. (\ref{a9})
and (\ref{a11}) we will deal with polydromic dreibein and gravitino fields,
which are still acceptable since the physical observables are bilinear
combination of them ( metrics and currents ), invariant under their 
monodromies.

\subsection{Mapping ( $X^a$, $\psi$ ) as mapping in superspace}

Supergravity has been noted firstly in field theory but then more
self-consistent mathematical formalisms have been developed
\cite{suma}-\cite{roge}, which make use of the concepts of extended
coordinates and supermanifolds. Since our method of solution is 
based on coordinate transformations, we will reformulate the problem
of finding a self-consistent solution for the metric and motion of
$N$-superparticles in terms of superspace transformations.   

At the end of the last paragraph we noted that, in a singular gauge, it
was possible to define the coupling with point sources by introducing
polydromic conditions on the fields around the particle sites. Here, we want
to express these as polydromic conditions on a mapping between 
supercoordinates, carrying cuts around a set of $N$-point sources:
\begin{eqnarray}
T = T ( t, z, \overline{z}, \theta, \overline{\theta} ) \nonumber \\
Z = Z ( t, z, \overline{z}, \theta, \overline{\theta} ) \nonumber \\
\Theta = \Theta ( t, z, \overline{z}, \theta, \overline{\theta} ).
\label{a12} \end{eqnarray}
To generate non-trivial solutions to the $(2+1)$-supergravity field equations, 
we must carefully choose specific cuts which are compatible with the 
invariance of the supermetric $ds^2$, and therefore related to
``isometries'' of the flat metric:
\beq ds^2 = {( dT + \frac{i}{2} ( \Theta d \overline{\Theta} +
\overline{\Theta} d \Theta ) )}^2 - ( dZ + \Theta d \Theta ) ( d
\overline{Z} + d \overline{\Theta} \overline{\Theta} ) - d\Theta
d\overline{\Theta} .\label{a13}\eeq

Let us note the presence of torsion in superspace that modifies the
ordinary differentials to be supersymmetric invariant differentials:

\begin{eqnarray}
dT \rightarrow dT  & + & \frac{i}{2} ( \Theta d \overline{\Theta} + 
\overline{\Theta} d \Theta ) \nonumber \\
dZ \rightarrow dZ & + & \Theta d \Theta .
\label{a14} \end{eqnarray}

Moreover we have to know how to relate the data of the superparticle with the
parameters of the corresponding monodromy in superspace.
In general, the superparticle has to be represented with an extended
set of coordinates $\xi^{M} = ( \xi^0 , \xi, \overline{\xi}, \xi^F,
\overline{\xi}^F )$. The relativistic action of a free superparticle 
has an internal fermionic symmetry that makes possible, in a particular
gauge, to choose the solution of the equations of motion in
such a way that the fermionic coordinate is constant:
\beq \xi^0 = \gamma t \ \ \ \  \xi(t) = V \xi^0 \ \ \ \ \xi^F (t) = 
\epsilon_0 .\label{a15} \eeq
To represent the data of a free superparticle, we must add to the constant
velocity $V$ the fermionic constant $\epsilon_0$. In the case of gravity
we have been able to represent the space-time of $N$ particles as a 
non trivial mapping between a Minkowskian coordinate system and a
physical one $X^a= X^a(x^b)$. The typical conical singularity of a
particle is described by a Lorentz cut in the $X^a$ flat coordinate
system, i.e. we impose that, when turning around the i-th particle site
$\xi_i$ in the physical coordinate system
$(z - \xi_i) \rightarrow e^{2\pi i} \ (z - \xi_i)$
\begin{eqnarray}
& \Delta T & \rightarrow ( a_i \overline{a}_i + b_i \overline{b}_i )
\Delta T 
+ \overline{a}_i b_i \Delta Z
+ a_i \overline{b}_i \Delta \overline{Z} \nonumber \\
& \Delta Z & \rightarrow 2 \overline{a}_i \overline{b}_i 
\Delta T + \overline{a}^2_i \Delta Z +
\overline{b}^2_i \Delta \overline{Z}, 
\label{a16} \end{eqnarray}
where $\Delta T = T - T(\xi_i)$, $\Delta Z = Z - Z(\xi_i)$ are
distances between a generic point and the particle site. 

We call them $DJH$ matching conditions; these make possible to
describe the scattering of point particles in a Minkowskian coordinate system,
a property that assures the integrability of the $N$-body problem.
In the Minkowskian coordinate system all particles move freely with
constant velocities $V_i$, ($Z(\xi_i) = V_i T(\xi_i) + B_i$ )
and the coefficients of the Lorentz cuts are
related to the particle data as follows \cite{ciaf}
\begin{eqnarray}
a_i & = & cos \frac{m_i}{2} + i \gamma_i sin \frac{m_i}{2}, \ \ \ 
\nonumber \\
b_i & = & - i \gamma_i \overline{V}_i  sin \frac{m_i}{2} \nonumber \\
V_i & \equiv & \pm \frac{P_i}{E_i}, \ \ \gamma_i = 
{( 1 - | V_i |^2 )}^{-\frac{1}{2}}, i = 1,2,...,n . 
\label{a17} \end{eqnarray}
Here the $P^a_i$ are the conserved Minkowskian momenta 
\beq P^a_i \ = \ ( E_i / P_i / {\bar P}_i ) \ = \ m \gamma_i \ 
( 1 / V_i / {\bar V}_i ) \ . \label{a18} \eeq

Is the cut of eq. (\ref{a16}) an isometry of the flat metric in
superspace ? Generally no, unless we modify also the variable $\Theta$
in the following way:
\begin{eqnarray}
 \Theta & \rightarrow & \overline{a}_i \Theta + i \overline{b}_i
\overline{\Theta} \nonumber \\
\overline{\Theta} & \rightarrow & a_i \overline{\Theta} - i b_i \Theta.
\label{a19} \end{eqnarray}
Then it is not difficult to show that the differentials $dT$, $dZ$,
and those supersymmetric invariant ( $dT + \frac{i}{2} ( \Theta d 
\overline{\Theta} + \overline{\Theta} d \Theta ), dZ + \Theta d \Theta$ )
satisfy the same Lorentz polydromy, and therefore in this case the
metric $ds^2$ remains invariant.

The last property implies that the Lorentz monodromy is compatible with the
supersymmetry transformations

\begin{eqnarray}
dT & \rightarrow & dT + 
\frac{i}{2} ( d\Theta \overline{\epsilon}_i +
d\overline{\Theta} \epsilon_i ) \nonumber \\
dZ & \rightarrow & dZ + d\Theta \epsilon_i \nonumber \\
\Theta & \rightarrow & \Theta + \epsilon_i ,
\label{a20} \end{eqnarray}
and the more general monodromy transformations, keeping $ds^2$
invariant and defining the coupling of superparticles to supergravity, are:

\begin{eqnarray}
\Delta T & \rightarrow & ( a_i \overline{a}_i + b_i \overline{b}_i ) 
( \Delta T + \frac{i}{2} (
\Theta \overline{\epsilon}_i + \overline{\Theta} \epsilon_i ) ) +
\overline{a}_i b_i ( \Delta Z + \Theta \epsilon_i ) + a_i \overline{b}_i 
( \Delta \overline{Z}
+ \overline{\epsilon}_i \overline{\Theta} ) \nonumber \\
\Delta Z & \rightarrow & 2 \overline{a}_i \overline{b}_i ( 
\Delta T + \frac{i}{2} ( \Theta
\overline{\epsilon}_i + \overline{\Theta} \epsilon_i ) ) + \overline{a}^2_i
( \Delta Z + \Theta \epsilon_i ) + \overline{b}^2_i ( \Delta \overline{Z} +
\overline{\epsilon}_i \overline{\Theta} ) \nonumber \\
\Theta & \rightarrow & \overline{a}_i ( \Theta + \epsilon_i ) + i 
\overline{b}_i ( \overline{\Theta} + \overline{\epsilon}_i ),
\label{a21} \end{eqnarray}
where as before $\Delta T = T - T(\xi_i), \ \Delta Z = Z -
Z(\xi_i)$ and $Z(\xi_i) - V_i T(\xi_i) =  B_i$. Let us note the following
remarkable property of these monodromies, i.e the fact that,
by introducing supersymmetric invariant distances 
\begin{eqnarray}
\Delta_S T & = & 
T - T(\xi_i) - \frac{i}{2} ( \Theta \overline{\Theta} (\xi_i) + 
\overline{\Theta} \Theta(\xi_i) ) \nonumber \\
\Delta_S Z & = & Z - Z(\xi_i) - \Theta \Theta(\xi_i) \nonumber \\
\Delta_S \Theta & = & \Theta - \Theta(\xi_i) ,
\label{a22} \end{eqnarray}
these can be recast in a more elegant form
\begin{eqnarray}
\Delta_S T & \rightarrow & ( a_i \overline{a}_i + b_i \overline{b}_i )
\Delta_S T + \overline{a}_i b_i \Delta_S Z
+ a_i \overline{b}_i \Delta_S \overline{Z} \nonumber \\
\Delta_S Z & \rightarrow & 2 \overline{a}_i \overline{b}_i 
\Delta_S T + \overline{a}^2_i \Delta_S Z +
\overline{b}^2_i \Delta_S \overline{Z} \nonumber \\
\Delta_S \Theta & \rightarrow & \overline{a}_i \Delta_S \Theta + i 
\overline{b}_i \Delta_S \overline{\Theta} ,
\label{a23} \end{eqnarray}
where $\Theta(\xi_i)$ are defined as fixed points of the
$\Theta$ monodromy.

To be more precise, the correspondence is not completely exact. In fact,
there is a missing piece $O(\epsilon \overline{\epsilon})$, between
eq. (\ref{a21}) and eq. (\ref{a23}). If we believe in the last
equation, which is manifestly supersymmetric \footnote{We also must
require that when $X^M = X^M (\xi_i)$ the monodromy reduces to $0
\rightarrow 0$.}, we have to correct eq. (\ref{a21}) as:
\begin{eqnarray}
\Delta T & \rightarrow & ( a_i \overline{a}_i + b_i \overline{b}_i ) 
( \Delta T + \frac{i}{2} (
\Theta \overline{\epsilon}_i + \overline{\Theta} \epsilon_i ) ) +
\overline{a}_i b_i ( \Delta Z + \Theta \epsilon_i ) + a_i \overline{b}_i 
( \Delta \overline{Z}
+ \overline{\epsilon}_i \overline{\Theta} ) \nonumber \\
& + & \frac{i}{2} (a-\overline{a}) \Theta(\xi_i)
\overline{\Theta}(\xi_i) \nonumber \\
\Delta Z & \rightarrow & 2 \overline{a}_i \overline{b}_i ( 
\Delta T + \frac{i}{2} ( \Theta
\overline{\epsilon}_i + \overline{\Theta} \epsilon_i ) ) + \overline{a}^2_i
( \Delta Z + \Theta \epsilon_i ) + \overline{b}^2_i ( \Delta \overline{Z} +
\overline{\epsilon}_i \overline{\Theta} ) +i \overline{b}
\Theta(\xi_i) \overline{\Theta} (\xi_i) \nonumber \\
\Theta & \rightarrow & \overline{a}_i ( \Theta + \epsilon_i ) + i 
\overline{b}_i ( \overline{\Theta} + \overline{\epsilon}_i ) . 
\label{a24} \end{eqnarray}
We have discovered a translation monodromy, not to be confused with
the typical translation monodromy of the spinning sources \cite{spin}.

We can also write down a more general case, in which a real spinning
source $S_i$ is present:
\begin{eqnarray}
 \Delta_S X^a \ & \rightarrow & \ L^a_{i \ b} \Delta_S X^b + \frac{S_i}{m_i}
P^a_i \nonumber \\
 \Delta_S \Theta \ & \rightarrow & \ \overline{a}_i \Delta_S \Theta + i
\overline{b}_i \Delta_S \overline{\Theta}.
\label{a25} \end{eqnarray}
We will see that this extra term is unavoidable in the explicit
solution and it will give rise to a logarithmic behaviour of the 
$X^M$ supercoordinates around the superparticle sites, breaking the manifest
supersymmetry of eq (\ref{a23}).

Finally, let us note that, starting from the knowledge of the
coordinate transformation in superspace, we can deduce the dreibein, 
which is polydrome only with respect to the Lorentz part of the monodromy

\begin{eqnarray}
e^a_\mu & = & ( \partial_\mu T + \frac{i}{2} ( \Theta \partial_\mu 
\overline{\Theta} + \overline{\Theta} \partial_\mu \Theta ),
\partial_\mu Z + \Theta \partial_\mu \Theta, \partial_\mu \overline{Z}
+ \partial_\mu{\overline \Theta} \overline{\Theta} ) \nonumber \\
\psi_\mu & = & \partial_\mu \Theta .
\label{a26} \end{eqnarray}

This parameterization of the solution makes explicit that the source
of torsion in the dreibein equation is solved by the presence of
bilinear terms in $\Theta$. From now on we will concentrate only on 
computing the supercoordinate transformation, which is enough to solve
the superparticle motion.

\subsection{Solution for the cuts}

In order to define the supermapping (\ref{a12}), we have to impose some gauge
condition. In gravity we have learned that it exists a gauge
condition for the metric  in which all the metric components 
propagate instantaneously \cite{bell} and that space-time can be
foliated in terms of space-like hypersurfaces as in the ADM formalism 
\cite{admf}. It can also  be defined in a first order formalism
as a $X^a = X^a(x^b)$ mapping that can be decomposed as a sum of
analytic and antianalytic functions \cite{ciaf}. 
We are going to extend the second definition of instantaneous gauge to
the case of supergravity, since searching a generalization in the
first order formalism, based on the dreibein, is simpler than looking
for a gauge condition for the metric in this theory. In fact we can
provide a realization of the cuts for the coordinates $T$, $Z$,
$\Theta$ starting from analytic functions

\begin{eqnarray}
T & = & T(t) + A_0 (z) + \overline{A_0}( \overline{z}) + O(\theta, 
\overline{\theta}) \nonumber \\
Z & = & A_1(z) + \overline{A_2} (\overline{z}) + O(\theta, \overline{\theta})
\nonumber \\
\Theta & = & \psi (z) + \overline{\chi} ( \overline{z} ) + O(\theta,
\overline{\theta}) .
\label{a27} \end{eqnarray}

Solving the $DJH$ matching conditions around the superparticles implies the
following conditions on the cuts of the analytic functions

\begin{eqnarray}
{A'}_0(z) & \rightarrow & ( a \overline{a} + b \overline{b} ) ( {A'}_0 (z) +
\frac{i}{2}
( {\psi'}(z) \overline{\epsilon} + {\chi'}(z) \epsilon )) + \overline{a} b (
{A'}_1(z) + {\psi'}(z) \epsilon ) + a \overline{b} ( {A'}_2(z) +
\overline{\epsilon} {\chi'}(z) ) \nonumber \\
{A'}_1(z) & \rightarrow & 2 \overline{a} \overline{b} ( {A'}_0 (z) +
\frac{i}{2} ( {\psi'}(z) \overline{\epsilon} + {\chi'}(z) \epsilon ) ) + 
\overline{a}^2 ( {A'}_1 (z) + {\psi'}(z) \epsilon ) + 
\overline{b}^2 ( {A'}_2 (z) + 
\overline{\epsilon} {\chi'}(z) ) \nonumber \\
{A'}_2 (z) & \rightarrow & 2 ab ( {A'}_0 (z) + \frac{i}{2} ( {\psi'}(z)
\overline{\epsilon} + {\chi'}(z) \epsilon ) ) + b^2 ( {A'}_1 (z) + {\psi'}(z)
\epsilon ) + a^2 ( {A'}_2 (z) + \overline{\epsilon} {\chi'}(z) ) \nonumber \\
\psi (z) & \rightarrow & \overline{a} ( \psi (z) + 
\frac{\epsilon}{2} ) +
i \overline{b} ( \chi + \frac{\overline{\epsilon}}{2} ) \nonumber \\
\chi (z) & \rightarrow & a ( \chi (z) + 
\frac{\overline{\epsilon}}{2} ) - i b (
\psi + \frac{\epsilon}{2} ) .
\label{a28} \end{eqnarray}

In the last two equations we made the hypothesis that the
supersymmetric translation of $\Theta$ is divided in equal parts
between $\psi$ and $\chi$. The monodromy problem (\ref{a28}) seems too
difficult to solve, as it is formulated. Our strategy will be to
define supersymmetric invariant analytic functions, which turn out to
be useful for reducing this system to a more manageable one.

In fact, being $( \psi', \chi' )$ invariants under supersymmetric cuts,
we can integrate those defining some new combinations which are again 
covariant under Lorentz transformations 

\begin{eqnarray}
A_0' & + & i ( \chi \psi' + \psi \chi' ) \rightarrow ( a \overline{a} +
b \overline{b} ) ( A_0' + i ( \chi \psi' + \psi \chi' )) +
\overline{a} b ( A_1' + 2 \psi \psi' ) + a \overline{b} ( A_2' + 2 \chi'
\chi ) \nonumber \\
A_1' & + & 2 \psi \psi' \rightarrow 2 \overline{a} \overline{b} ( A_0'
+ i ( \chi \psi' + \psi \chi' )) + \overline{a}^2 ( A_1' + 2 \psi \psi'
) + \overline{b}^2 ( A_2' + 2 \chi' \chi ) \nonumber \\
A_2' & + & 2 \chi' \chi \rightarrow 2 a b ( A_0' + i ( \chi \psi' +
\psi \chi' )) + b^2 ( A_1' + 2 \psi \psi' ) + a^2 ( A_2' + 2 \chi' \chi ). 
\label{a29} \end{eqnarray}

Thus, we can identify these analytic functions with the ones
introduced in ($2+1$) gravity

\begin{eqnarray}
{A'}_0 (z) & + & i ( \chi (z) \psi' (z) + \psi (z) \chi' (z) )  
=  \frac{N(z)}{f'(z)} f(z)
\nonumber \\
{A'}_1 (z) & + & 2 \psi (z) \psi' (z) = \frac{N(z)}{f'(z)} \nonumber \\
{A'}_2 (z) & + & 2 \chi' (z) \chi (z) = \frac{N(z)}{f'(z)} f^2 (z) .
\label{a30} \end{eqnarray}

The function $N(z,t)$ is a meromorphic function with poles at the singularities
and for the two-body problem was found to be \cite{ciaf}:

\beq N(z,t) = C \xi^{1-\frac{{\cal M}}{2\pi}}
\frac{1}{(z-\xi_1)(\xi_2-z)} \ \ \ \ \ \xi = \xi_2 - \xi_1 , 
\label{a31} \eeq
and the function $f(z,t)$ can be exactly given for
the two-body problem (see Appendix), and perturbatively for the
$N$-body problem. In the two body case the function has branch points
at $\zeta = 0$ and $\zeta = 1$ ( and
$\zeta = \infty$ ), around which it has to transform projectively as

\beq f(z,t) \rightarrow \frac{a_i f(z,t) + b_i}{\overline{b}_i f(z,t)
+ \overline{a}_i} . \label{a32} \eeq

The $f$ function is obtained as a ratio of two independent solutions
$y_1$ and $y_2$ of the Fuchsian equation in the adimensional variable
$\zeta = ( z-\xi_1 ) / (\xi_2 - \xi_1 )$
\beq 
y'' + \frac{1}{4} \left( \frac{1-\mu_1^2}{\zeta^2} + 
\frac{1-\mu_2^2}{(1-\zeta)^2} +
\frac{1-\mu_1^2-\mu_2^2+\mu_{\infty}^2}{\zeta ( 1- \zeta )} \right) y
= 0 .
\label{a33} \eeq
The indices $\mu_1$, $\mu_2$ and $\mu_{\infty}$, appearing in eq. (\ref{a33}),
are related to the masses $m_1$, $m_2$ and the physical
invariant mass $\cal{M}$ as follows
\begin{eqnarray}
\mu_1 & = & \frac{m_1}{2\pi} , \ \ \ \ \ \mu_2 = \frac{m_2}{2\pi}
\nonumber \\
\mu_{\infty} & = & \left( \frac{\cal M}{2\pi} - 1 \right),
\label{a34} \end{eqnarray}
where $\cal M$, corresponding to the topological invariant $Tr ( L_1
L_2 ) $, is given by the Carlip's formula \cite{carl} 
\beq cos \frac{\cal M}{2} = cos \frac{m_1}{2} cos \frac{m_2}{2} - 
\frac{P_1 \cdot P_2}{m_1 m_2} sin \frac{m_1}{2} sin \frac{m_2}{2}.
\label{a35} \eeq
In the case of $N$-body the function $f(z,t)$ is determined again
by the same procedure, however the Schwarzian in this case contains not 
only poles at the $(N+1)$ physical singularities but also poles at $N-2$
apparent singularities, whose motion, as a function of the
physical poles, is constrained by the requirement of isomonodromy
of the monodromy matrices, as discussed in \cite{ciaf}-\cite{well}-\cite{isom}.

The non-triviality of supergravity is hidden in finding a solution for
the gravitino fields $\psi$ and $\chi$, 
which is simpler than the starting problem in
eq. (\ref{a28}), since we have to look for only a spin $\frac{1}{2}$ 
representation of the monodromies. It was
exactly this reduction that has permitted us to write the non
perturbative solution for the two body problem.
In order to determine $\psi$ and $\chi$ we have to solve the monodromy
conditions

\begin{eqnarray}
 \psi & \rightarrow & \overline{a} ( \psi + \frac{\epsilon}{2} ) + i
\overline{b} ( \chi + \frac{\overline{\epsilon}}{2} ) \nonumber \\
 \chi & \rightarrow & a ( \chi + \frac{\overline{\epsilon}}{2} ) - i b
( \psi + \frac{\epsilon}{2} ) .
\label{a36} \end{eqnarray}

The terms of order $\theta$ in the $X$-supermapping (\ref{a27}) can be
built from combinations of the functions of order zero in $\theta$ that
respect the Lorentz and supersymmetric cuts, from which we obtain the 
complete solution in superspace

\begin{eqnarray}
T & = & T(t) + A_0(z) + \overline{A}_0 ( \overline{z}) + \theta  \left[
i k_0 \chi \sqrt{{A'}_1 + 2 \psi \psi' } + k_0 
\psi \sqrt{{A'}_2 + 2 \chi' \chi } 
\right] \nonumber \\
& + &  \left[ - i {\overline k}_0 \overline{\chi} \sqrt{ \overline{A'}_1 + 2
\overline{\psi}' \overline{\psi}} + {\overline k}_0 \overline{\psi}
\sqrt{\overline{A'}_2 + 2 \overline{\chi}
\overline{\chi}' } \right] \overline{\theta} \nonumber \\
& -i & \theta \overline{\theta} 
\left[ h_0 ( \psi \chi' + \chi \psi' ) + {\overline h}_0
( \overline{\psi} \overline{\chi'} + \overline{\chi} \overline{\psi'}
) \right] \nonumber \\
Z & = & A_1(z) + \overline{A_2} ( \overline{z} ) + 2 \theta k_0 
\psi \sqrt{{A'}_1
+ 2 \psi \psi'} - 2i {\overline k}_0 
\overline{\chi} \sqrt{\overline{A'}_2 + 2
\overline{\chi} \overline{\chi}' } \overline{\theta} \nonumber \\
& - & 2 \theta \overline{\theta}  
\left[ h_0 \psi \psi' + {\overline h}_0 \overline{\chi}
\overline{\chi'} \right] \nonumber \\
\Theta & = & \psi (z) + \overline{\chi} ( \overline{z} ) + \theta
k_0 \sqrt{ {A'}_1 + 2 \psi \psi'} + i 
\overline{\theta} {\overline k}_0 \sqrt{ \overline{A'}_2 +
2 \overline{\chi} \overline{\chi}' } \nonumber \\
& + & \theta \overline{\theta} ( h_0 \psi' + 
{\overline h}_0 \overline{\chi'} ) ,
\label{a37} \end{eqnarray}
where $k_0$ and $h_0$ are arbitrary constant which are left
undetermined by the monodromies. In the following we will set, for
simplicity, $k_0=1$ which is equivalent to a rescaling of $\theta
\rightarrow \theta / k_0$. To derive eq. (\ref{a37}) we have been 
inspired by the form of superanalytic functions 
\footnote{We have avoided to consider terms 
which do not respect the condition that, when the source is decoupled
($\epsilon = m = 0$), the supermapping reduces to the identity. 
This requirement completes our gauge choice.}.

Naively, this definition of the $X$-supermapping can be shown to
satisfy the rule (\ref{a23}). In fact, the combinations $\Delta_S X^a$
can be written as follows:

\begin{eqnarray}
\Delta_S T & = & \int_{\xi_i}^z \ \left( \frac{N}{f'} f 
- i ( \Delta_S \psi \chi' + \Delta_S \chi \psi' )  \right) dz + 
 \theta  \left[ i \Delta_S \chi \sqrt{\frac{N}{f'}} + 
\Delta_S \psi \sqrt{\frac{N}{f'} f} 
\right] + \nonumber \\
& -i & \theta \overline{\theta} 
 h_0 ( \Delta_S \psi \chi' + \Delta_S \chi \psi' ) + (h.c.)
\nonumber \\
\Delta_S Z & = &  \int_{{\xi}_i}^z \left( \frac{N}{f'} - 2 \Delta_S \psi
\psi' \right) dz + \int_{\overline{\xi}_i}^{\overline{z}} \left( 
\frac{\overline{N}}{\overline{f}'} \overline{f}
- 2 \Delta_S \overline{\chi}
\overline{\chi}' ( \overline{z} ) \right) d\overline{z} + 2 \theta  
\Delta_S \psi \sqrt{\frac{N}{f'}}  - \nonumber \\
& - & 2i  
\Delta_S \overline{\chi} \sqrt{\frac{\overline{N}}{\overline{f'}}\overline{f}
} \overline{\theta} - 2 \theta \overline{\theta}  
\left[ h_0 \Delta_S \psi \psi' + {\overline h}_0 \Delta_S \overline{\chi}
\overline{\chi'} \right] \nonumber \\
\Delta_S \Theta & = & \Delta_S \psi (z) + \Delta_S \overline{\chi} 
( \overline{z} ) + \theta \sqrt{\frac{N}{f'}} + i \overline{\theta} 
\sqrt{\frac{\overline{N}}{\overline{f'}}
\overline{f}} +  \theta \overline{\theta} ( h_0 \psi' + 
{\overline h}_0 \overline{\chi'} ) .
\label{a38} \end{eqnarray}
In this formula we have introduced the notations 
$\Delta_S \psi = \psi (z) - \psi
(\xi_i)$ and $\Delta_S \chi = \chi (z) - \chi (\xi_i)$, where 
$\psi (\xi_i)$ and $\chi (\xi_i)$
are defined as fixed points of the $\psi$ and $\chi$ monodromies. Since 
$\Delta_S \psi $ and $\Delta_S \chi$ transform covariantly under
Lorentz transformations without translation terms, 
the whole mapping formally obeys eq. (\ref{a23}), unless the
integration of $1/z$ terms produces logarithms spoiling supersymmetry
covariance.
It will turn out, by analyzing in detail the one-body problem, that
the explicit solution breaks this covariance.

\section{Two-superbody problem}

In this section we will take advantage of this explicit
parameterization in order to have detailed informations about
superparticle dynamics. Let us start with the one-body static problem
and notice that, at the singularity, the value of $\Theta$ is fixed by
the monodromy conditions
\beq \Theta (0) = \frac{\epsilon}{a-1} . \label{b1}\eeq
Since from gravity we know that $f = 0$, and $N / f' = z^{-\mu}$, it
remains to look for a solution to the system (\ref{a36}), of the type
\beq
\psi' = \delta_1  z^{-\frac{\mu}{2}-1} \ \ \ \
\chi' =  \delta_2  z^{\frac{\mu}{2}-1}.
\label{b2} \eeq
Requiring that $\psi  \rightarrow \overline{a} ( \psi +
\frac{\epsilon}{2} )$ and $ \chi \rightarrow a ( \chi + 
\frac{\overline{\epsilon}}{2} )$ fixes only the boundary values
$\psi (0) = \overline{\chi} (0) = \Theta (0) / 2 $,
and leaves $\delta_1$ and $\delta_2$ undetermined:
\begin{eqnarray}
\psi (z) & = & \psi (0) - \frac{2}{\mu} \delta_1 
z^{-\frac{\mu}{2}} \nonumber \\
\chi (z) & = & \chi (0) + \frac{2}{\mu} \delta_2
z^{\frac{\mu}{2}}.
\label{b3} \end{eqnarray}

Substituting these expressions into eq. (\ref{a37}), we can compute
the total supercoordinate transformation as:
\begin{eqnarray}
T( z, \overline{z}, t) & = &  T_0(t) - i \int^z_0 \ dz \ ( \psi \chi' +
\chi \psi' ) + \theta i \chi \sqrt{\frac{N}{f'}} 
- i \theta \overline{\theta} h_0 ( \psi \chi' + \chi \psi' ) 
+ (h.c.) = \nonumber \\
& = & T_0(t)  + \frac{2i}{\mu} \chi (0) \delta_1 z^{-\frac{\mu}{2}} -
\frac{2i}{\mu} \psi (0) \delta_2 z^{\frac{\mu}{2}} + \frac{4i}{\mu} 
\delta_1 \delta_2 log z + \nonumber \\
& - & i \chi (0) \theta z^{-\frac{\mu}{2}} + \frac{2i}{\mu} \theta
\delta_2 - i \theta \overline{\theta} h_0 \left( \psi(0) \delta_2 
z^{\frac{\mu}{2}-1} + \chi(0) \delta_1 z^{-\frac{\mu}{2}-1} \right)
+ (h.c.) 
\nonumber \\
Z(z) & = & \frac{z^{1-\mu}}{1-\mu} + \frac{4}{\mu} \psi(0) \delta_1 
z^{-\frac{\mu}{2}} - \frac{4}{\mu} \overline{\chi} (0)
\overline{\delta}_2 \overline{z}^{\frac{\mu}{2}} + \nonumber \\
& + & 2 \theta \psi(0) z^{-\frac{\mu}{2}} - \frac{4}{\mu} \theta
\delta_1 z^{-\mu} - 2 \theta \overline{\theta} 
\left( h_0 \psi(0) \delta_1 
z^{-\frac{\mu}{2}-1} + \overline{h}_0 \overline{\chi} (0)
\overline{\delta}_2 \overline{z}^{\frac{\mu}{2}-1} \right)
\nonumber \\
\Theta(z) & = & \Theta(0) - \frac{2}{\mu} \delta_1 z^{-\frac{\mu}{2}}
+ \frac{2}{\mu} \overline{\delta}_2  
\overline{z}^{\frac{\mu}{2}} + \theta z^{-\frac{\mu}{2}}
\nonumber \\
& + & \theta \overline{\theta} \left( h_0 \delta_1 z^{-\frac{\mu}{2}-1} + 
\overline{h}_0 \overline{\delta}_2 \overline{z}^{\frac{\mu}{2}-1} 
\right). \label{b4} \end{eqnarray}

Around zero, the logarithmic contribution gives rise to an additional
translation monodromy for T, which has no relation with the
translation of (eq. (\ref{a24}) ) since it comes from the
integration of an unwanted $1/z$ term,  
\beq 
\Delta_S T \rightarrow \Delta_S T -\frac{16}{\mu} Re (\delta_1 \delta_2) , 
\label{b5} \eeq
and instead reveals the presence of a spin source $S \sim Re( 
\delta_1 \delta_2 )
$, as in eq. (\ref{a25}). Since this definition obeys the condition 
$S^2 = 0$, it is possible to avoid the $CTC$ problem, which was encountered
in the solution of spinning particles, with the spin variable treated
as a pure classical variable. We interpret it not as an external
spinning source but as a spontaneous violation of supersymmetry in
presence of a conical singularity.

Since at the singularity the value of $\Theta(0)$ is trivially
reproduced, apparently there is no condition to constraint the fermionic
trajectory. Here there is a difference from the bosonic case, where
requiring that the $X^a$ mapping has a specific value at the particle
site $X^a(\xi_i)$ was enough to determine the general first integral of the
geodesic equations. However there is a singularity of the
mapping at the particle site, as in the case of spinning particles
coupled to gravity, $\psi (z) \sim z^{-\frac{\mu}{2}}$. Moreover, in this
case the superspace geometry can
help to give an interesting interpretation to it. In fact, to
understand better the role of this divergence we must add the
contribution in $\theta$  which dresses the divergent part in $\psi (z)$

\beq \Theta = \left( \theta - \frac{2}{\mu} \delta_1 \right) 
z^{-\frac{\mu}{2}} + \theta \overline{\theta}  
\left( h_0 \delta_1 z^{-\frac{\mu}{2}-1} + \overline{h}_0 \overline{\delta}_2 
{\overline z}^{\frac{\mu}{2}-1} \right)  ... \label{b6} \eeq
To have a well defined theory at the classical level, we require that
the singular behaviour disappears, due to the contribution of $\theta (0)$:

\beq \theta (0) = \xi^F =  \frac{2}{\mu} \delta_1 . \label{b7} \eeq

This is the obvious solution to the spinor part of the 
supergeodesics equations \cite{suma} for a static superparticle.
The spinor coordinate $\xi^F$ is ill-defined when the mass 
$m \rightarrow 0$ but this is correct because a massless particle 
at rest , carrying an additional intrinsic degree of freedom ( the
constant spinor $\epsilon$ ), is not a self-consistent concept.

The contribution at the singularity of the type $z^{-\frac{\mu}{2}}$
in the bosonic coordinates $T, Z, \overline{Z}$ is again
cancelled due to eq. (\ref{b7}), whatever the value of $\delta_1$ is. 
In fact for the T coordinate we obtain the singular contribution:
\beq T|_{\rm sing} \ \sim \ - i \chi(0) \left( \theta - \frac{2}{\mu} 
\delta_1 \right) z^{-\frac{\mu}{2}} ,   \label{b8} \eeq
and similarly for the $Z$ coordinate 
\beq Z|_{\rm sing} \ \sim \ 2 \left( \theta - \frac{2}{\mu}
\delta_1 \right) \psi(0) z^{-\frac{\mu}{2}}.   \label{b9} \eeq

The $\theta \overline{\theta}$ terms induce other divergences 
$z^{\pm\frac{\mu}{2}-1}$, which cannot be mixed with
$z^{-\frac{\mu}{2}}$. To get a cancellation of them we should require
that $\delta_2$ is a linear combination of $\delta_1$ and 
$\overline{\delta}_1$. It is of course possible to restrict the
arbitrary parameters $\delta_1$ and $\delta_2$ to do so, but, for more
than one particle, this requirement would lead to an involved
solution, carrying many parameters. To keep the discussion as more
concise as possible, we will set in the following the arbitrary
parameter $h_0 = 0$, since then a minimal solution ( but not trivial )
is shown to be consistent at all orders.

In conclusion, while in the bosonic case the finite values $X^a(x_i)$
are responsible for the $\xi (t)$ bosonic trajectory, in the fermionic
case we find that the divergent behaviour of the $X^M$ mapping around
the particle site defines the $\xi^F (t)$ fermionic
trajectory and the finite values $\Theta(\xi_i)$ 
are trivially reobtained by the explicit solution.

The two-body static case is similarly obtained by the monodromies :

\begin{eqnarray}
\psi \ {\buildrel{1} \over \rightarrow} \ \overline{a}_1 ( \psi +
\frac{\epsilon_1}{2} ) & \ & \chi \
{\buildrel{1} \over \rightarrow} \ a_1 ( \chi +
\frac{\overline{\epsilon_1}}{2} ) \nonumber \\
\psi \ {\buildrel{1} \over \rightarrow} \ \overline{a}_2 ( \psi +
\frac{\epsilon_2}{2} ) & \ & \chi \
{\buildrel{2} \over \rightarrow} \ a_2 ( \chi +
\frac{\overline{\epsilon_2}}{2} ) .
\label{b10} \end{eqnarray}

Let us suppose to have two superparticles at rest ( $z= \xi_i$  ), which
can be mapped by introducing the adimensional variable $\zeta = 
( z-\xi_1 ) / ( \xi_2-\xi_1 ) $, into the points $\zeta =0$ and $\zeta =
1$. At the singularity, the value of $\Theta$ is fixed by the monodromy
\beq \Theta (0) = 2\psi(0) = 2 \overline{\chi} (0) = 
\frac{\epsilon_1}{a_1-1}  \ \
\ \ \ \ \Theta (1) = 2\psi(1) = 2 \overline{\chi} (1) =
\frac{\epsilon_2}{a_2-1}.
\label{b11} \eeq

Let us look for a solution of the type:

\begin{eqnarray} 
\psi' & = & \left[ \frac{\delta_1}{\zeta} + 
\frac{\sigma_1}{1-\zeta} \right] {\zeta}^{- \frac{\mu_1}{2}}
{( 1- \zeta )}^{-\frac{\mu_2}{2}} \ \nonumber \\
\chi' & = &  \left[ \frac{\delta_2}{\zeta} + 
\frac{\sigma_2}{1-\zeta} \right]
{ \zeta }^{\frac{\mu_1}{2}}
{(1-\zeta)}^{\frac{\mu_2}{2}} .  
\label{b12} \end{eqnarray}

The condition which determines the unknowns $\delta_i, \sigma_i (i=1,2) $ are 
the fermionic distances, invariant under constant fermionic
translations of the $\Theta$-variable,
\begin{eqnarray} \psi (1) & - & \psi (0) = \int^1_0 \ d\zeta \ \psi'
= -2 \frac{\Gamma( 1- \frac{\mu_1}{2} ) \Gamma( 1- \frac{\mu_2}{2} )}{
\Gamma( 1- \frac{\mu_1}{2} -\frac{\mu_2}{2})} \left[
\frac{\delta_1}{\mu_1} + \frac{\sigma_1}{\mu_2} \right]   
\nonumber \\ \chi (1) & - & \chi (0) = \int^1_0 \ d\zeta \ \chi' 
= 2 \frac{\Gamma( 1 +\frac{\mu_1}{2} ) \Gamma( 1 +\frac{\mu_2}{2} )}{
\Gamma( 1 +\frac{\mu_1}{2} +\frac{\mu_2}{2})} \left[
\frac{\delta_2}{\mu_1} + \frac{\sigma_2}{\mu_2} \right] 
.\label{b13} \end{eqnarray}
The monodromy conditions for $\psi$ and $\chi$ leave always two free
parameters, in the general $N$-body case too. We can of course
determine the solution by adding some ``physical requirements'', but
we prefer to take into account only the necessary ones, such as the
monodromies. In any case this formalism can be generalized to support
more constraints on the anticommuting variables, by increasing the
number of parameters. 

As in the one-body problem, unavoidable logarithmic terms appear as follows
\begin{eqnarray}
T & \buildrel {\zeta \sim 0} \over \sim & \frac{4i}{\mu_1} \delta_1
\delta_2 log \zeta + (h.c.) \quad \quad \Rightarrow 
\Delta_S T \ {\buildrel{1} \over \rightarrow}  \ 
\Delta_S T - \frac{16}{\mu_1} Re (\delta_1 \delta_2)  \nonumber \\
& \buildrel {\zeta \sim 1} \over \sim &  \frac{4i}{\mu_2} \sigma_1
\sigma_2  log (\zeta-1) + (h.c.) \quad \quad \Rightarrow 
\Delta_S T \ {\buildrel{2} \over \rightarrow} \ 
\Delta_S T - \frac{16}{\mu_2} Re (\sigma_1 \sigma_2), \nonumber \\ & & 
\label{b14} \end{eqnarray}
revealing the presence of a spin $S_1 \sim Re(\delta_1 \delta_2)$ at $\zeta=0$
and $S_2 \sim Re(\sigma_1 \sigma_2)$ at $\zeta=1$.

Moreover there is a singular behaviour of the field $\psi$, which is
unphysical at the classical level. As in the one body case,
we obtain a cancellation of the divergence at the particle
site by adding the contribution in $\theta$:

\beq \Theta = \psi (z) + \overline{\chi} ( \overline{z}) +
\theta \sqrt{\frac{N}{f'}} + O ( \theta \overline{\theta} ) 
\label{b15} . \eeq

In the static case $f=0$, $\sqrt{\frac{N}{f'}} =
(z-\xi_1)^{-\frac{\mu_1}{2}} (\xi_2-z)^{-\frac{\mu_2}{2}}$,
therefore the $\theta$ term dresses 
the divergent part in $\psi_0 (z)$

\beq \Theta|_{\zeta \simeq 0} = \xi^{-\frac{\mu_2}{2}}
\left( \theta + \frac{2}{\mu_1} \delta_1  \xi^{\frac{\mu_1+\mu_2}{2}}
\right) ( z - \xi_1 )^{-\frac{\mu_1}{2}} \label{b16} \eeq
and defines the fermionic coordinates:

\beq \theta (0) = \xi^F_1 = - \frac{2}{\mu_1} \delta_1  
\xi^{\frac{\mu_1+\mu_2}{2}} \ \ \ \ \
\theta (1) = \xi^F_2 = \frac{2}{\mu_2} \sigma_1 
\xi^{\frac{\mu_1+\mu_2}{2}} .
\label{b17} \eeq

It is interesting to observe that the difference $\xi^F_2 - \xi^F_1$
is physically meaningful
\beq \theta (1) - \theta (0) = - \frac{\Gamma ( 1 - \frac{\mu_1}{2} -
\frac{\mu_2}{2} )}{\Gamma (1- \frac{\mu_1}{2}) \Gamma
(1-\frac{\mu_2}{2})} \xi^{\frac{\mu_1+\mu_2}{2}} ( \psi(1) - \psi(0) ),
\label{b18} \eeq
being dependent only on the constant of motions. Therefore the
aforementioned arbitrariness could be related to the choice of a 
fermionic center of mass, while the relative coordinate is unambiguous.

Let us check how the divergent terms $(z-\xi_i)^{-\frac{\mu_i}{2}}$
are cancelled by the $\theta$ contributions in the bosonic $X^a$
mappings. For example the divergence of the type 
$(z-\xi_1)^{-\frac{\mu_1}{2}}$ in the $T$ and $Z$-variables can be
written as follows
\begin{eqnarray}
T \ {\buildrel{1} \over \sim} \ 
-i \chi (0) \left( \psi + \theta \sqrt{\frac{N}{f'}} \right)  \quad
\quad \quad  Z \ {\buildrel{1} \over \sim} \ 
-2 \psi (0) \left( \psi + \theta \sqrt{\frac{N}{f'}} \right) ,
\label{b19} \end{eqnarray}
since $\psi$ and $\chi$ can be considered constant. Again the
cancellation of the fermionic residue at the singularity leads to
equation (\ref{b17}). 

The bosonic part of the geodetic equations is equivalent to impose
$Z(\xi_2) - Z(\xi_1) = B_2 - B_1$ and is of course trivial since 
only determines the distance $\xi$ in terms of the bosonic constant of 
motion $B_2-B_1$.

\subsection{Non-perturbative solution}

There are many non perturbative informations that can be extracted
without doing any calculations, in practice. By using eq. (\ref{a21}) 
the value of $\Theta$ at the particle site is determined as a fixed
point of the $\Theta$ monodromy as
\beq
\Theta_i = 2 \psi ( \xi_i ) = 2 \overline{\chi} ( \xi_i ) = 
\frac{(\overline{a}_i - 1) \epsilon_i + i \overline{b}_i 
\overline{\epsilon}_i }{2 - a_i - \overline{a}_i } . \label{b20}
\eeq
Moreover we note that the $\theta$-terms cannot give any 
finite contribution at the particle sites in the $\Theta$-mapping 
(\ref{a37}). Therefore, the conclusion that the fermionic
geodetic cannot be obtained by means of finite terms is simply a
corollary of the monodromy properties.

Without solving exactly the fermionic fields, it is possible to
discuss in full detail the fermionic geodetic equation, since the
monodromies tell us enough information to constraint the
behaviour of these fields around the superparticles. 

In particular, there are two leading behaviours $\zeta^{\pm \frac{\mu_1}{2}}$
around the superparticles and their coefficients are constrained in the
following way

\begin{eqnarray}
\psi - \psi(0) & \simeq & \Delta_1 \zeta^{-\frac{\mu_1}{2}} + i \overline{f}
(0) \Delta_2 \zeta^{\frac{\mu_1}{2}} + O(\zeta^{1-\frac{\mu_1}{2}}) +  
O(\zeta^{1+\frac{\mu_1}{2}}) \nonumber \\
\chi- \chi(0) & \simeq & -i f(0) \Delta_1 
\zeta^{-\frac{\mu_1}{2}} + \Delta_2 
\zeta^{\frac{\mu_1}{2}} + O(\zeta^{1-\frac{\mu_1}{2}}) +  
O(\zeta^{1+\frac{\mu_1}{2}}) \nonumber \\
f(0) & = & \frac{\gamma_1 \overline{V}_1}{1+\gamma_1} .
\label{b21} \end{eqnarray}

The explicit solution will lead to know exactly how the $\Delta_i$
coefficients are related to the fermionic constants of motion
$\epsilon_i$ (see later on). 

By using eq. (\ref{b21}) we can control all the divergences of the
$X^M$ mapping in the interacting two-body problem, and also in the general
$N$-body problem. Let us first describe the logarithmic cut in the
$X^a$ variables, around particle \#1 at $\zeta = 0$. By inserting the 
development of the solution near the particle site, the $T$ and $Z$ 
variables contain logarithmic cuts

\begin{eqnarray}
T & \sim & - i \int^z \ dz \ ( \chi \psi' + \psi \chi' ) + (h.c.)
= i \mu_1 \Delta_2 \Delta_1 ( 1 + f(0) \overline{f}(0) ) \ 
log \zeta + (h.c.)
\nonumber \\
Z & \sim & - 2 \int^{\zeta} \ d{\zeta} \ \psi \psi' 
- 2 \int^{\overline \zeta} \ d{\zeta} \ 
{\overline\chi} {\overline\chi}'
= 2 i \mu_1 \Delta_2 \Delta_1
\overline{f} (0) \ log \zeta - 2 i \mu_1 \overline{\Delta}_1
\overline{\Delta}_2 \overline{f} (0)  \
log \overline{\zeta}. \nonumber \\
& & \label{b22} \end{eqnarray}
Then we can conclude that the combination $Z-V_1T$ doesn't contain 
any logarithmic term around particle \#1 because :

\beq \frac{2 \overline{f} (0)}{ 1 + f(0) \overline{f} (0)} = V_1 . 
\label{b23} \eeq
This property will be very useful to discuss the bosonic part of the
geodetic equations.

To discuss the fermionic geodesic equations, it is enough to develop
the $\theta$ terms around particle \#1, obtaining that 

\begin{eqnarray}
\sqrt{\frac{N}{f'}} & \simeq & C \xi^{-\frac{{\cal M}}{4\pi}} 
\zeta^{-\frac{\mu_1}{2}} ( 1 + \overline{f} (0) \widetilde{C} \zeta^{\mu_1} ) 
+ O ( \zeta^{1-\frac{\mu_1}{2}} ) + O ( \zeta^{1+\frac{\mu_1}{2}} )
\nonumber \\
\sqrt{\frac{N}{f'}} f & \simeq & C \xi^{-\frac{{\cal M}}{4\pi}} 
\zeta^{-\frac{\mu_1}{2}} ( f(0) + \widetilde{C} \zeta^{\mu_1} )
+ O ( \zeta^{1-\frac{\mu_1}{2}} ) + O ( \zeta^{1+\frac{\mu_1}{2}} ) .
\label{b24} \end{eqnarray}

Having fixed the singular behaviour of the fields, it is rather clear
how to compute the fermionic geodesic. For example, from the
$\Theta$-mapping we find that 

\beq \Theta \sim \zeta^{-\frac{\mu_1}{2}} [ \Delta_1 + \theta C
\xi^{-\frac{{\cal M}}{4\pi}} ] + i \overline{f} (0) 
\overline{\zeta}^{-\frac{\mu_1}{2}} [ \overline{\Delta}_1 +
\overline{\theta} \overline{C} \overline{\xi}^{-\frac{{\cal M}}{4\pi}}
] , \label{b25} \eeq

which is solved by 
\beq \theta (0) = \xi^F_1 = 
- \frac{\Delta_1}{C}\xi^{\frac{{\cal M}}{4\pi}} .
\label{b26} \eeq
For the $Z$-mapping and $T$-mapping we find:

\begin{eqnarray}
Z & \sim & - 2 \psi(0) \zeta^{-\frac{\mu_1}{2}} [ \Delta_1 + \theta C
\xi^{-\frac{{\cal M}}{4\pi}} ] - 2i \overline{f} (0) \overline{\chi} (0) [
\overline{\Delta}_1 + \overline{\theta} \overline{C}
\overline{\xi}^{-\frac{{\cal M}}{4\pi}} ] 
{\overline{\zeta}}^{-\frac{\mu_1}{2}} \nonumber \\
T & \sim & - ( i \chi(0) + \psi(0) f(0) ) [ \Delta_1 + \theta C 
\xi^{-\frac{{\cal M}}{4\pi}} ] \zeta^{-\frac{\mu_1}{2}} + (h.c.),
\label{b27} \end{eqnarray}
and the additional constraint

\beq \theta(0) \psi|_{\rm sing} = \theta(0) \chi|_{\rm sing} = 0,
\label{b28} \eeq
where $\psi|_{\rm sing}$ denotes the singular part of $\psi$ around
the superparticle site. All of them are solved by a single equation 
(\ref{b26}). Therefore we have a non-perturbative proof that the
cancellation of divergences is valid at all orders.

We are ready to describe the non-perturbative solution. This will be
built in terms of hypergeometric functions, as in the case of
gravity. Let us define the following basis of the hypergeometric
equation
\begin{eqnarray}
y_{+} (0) & = & \zeta^{\frac{\mu_1}{2}}
{(1-\zeta)}^{-\frac{\mu_2}{2}}
\widetilde{F} ( a', b', c' ; \zeta ) \nonumber \\
y_{-} (0) & = & \zeta^{-\frac{\mu_1}{2}}
{(1-\zeta)}^{-\frac{\mu_2}{2}}
\widetilde{F} ( a, b, c ; \zeta ) , \label{b29} 
\end{eqnarray}
where we use the notation
\begin{eqnarray}
\widetilde{F} (a, b, c ; \zeta ) & = & \frac{\Gamma (a)
\Gamma (b)}{\Gamma (c)} \ F ( a, b, c ; \zeta ) . \label{b30}
\end{eqnarray} 
To match, in the static limit, the $\psi$ function, the correct
choice for the coefficients $a$, $b$, and $c$ turns out to be, as in
the gravity case,
\begin{eqnarray}
a & = & \frac{1}{2} \left( \frac{{\cal M}}{2\pi} -\mu_1 - \mu_2
\right) \nonumber \\
b & = & 1 - \frac{1}{2} \left( \frac{{\cal M}}{2\pi} +\mu_1 +\mu_2
\right) \nonumber \\
c & = & 1 -\mu_1  , \label{b31}
\end{eqnarray}
and  $a' = a - c + 1 = a(-\mu_1), b' = b - c + 
1 = b(-\mu_1), c' = 2 
- c = c(-\mu_1)$. The basis,
defined by eqs. (\ref{b29}) and (\ref{b31}) is also solution of the 
Fuchsian eq. (\ref{a33}) ( apart from an overall redefinition
of $y$ by power factors ) , with $\mu_{\infty} = \frac{{\cal M}}{2\pi} - 1$.

In the case of supergravity it is needed another basis, which we call 
$\tilde{y}_{\pm} (0)$, to match, in the static limit, the $\chi$ function.
It is similar to the above one with all the masses reversed in sign
and it is given by
\begin{eqnarray}
\tilde{y}_{+} (0) & = & \zeta^{-\frac{\mu_1}{2}}
{(1-\zeta)}^{\frac{\mu_2}{2}}
\widetilde{F} ( \tilde{a}', \tilde{b}', \tilde{c}' ; \zeta ) \nonumber \\
\tilde{y}_{-} (0) & = & \zeta^{\frac{\mu_1}{2}}
{(1-\zeta)}^{\frac{\mu_2}{2}}
\widetilde{F} ( \tilde{a}, \tilde{b}, \tilde{c} ; \zeta ) , \label{b32}
\end{eqnarray}
and 
\begin{eqnarray}
\tilde{a} & = & - \frac{1}{2} \left( \frac{{\cal M}}{2\pi} -\mu_1 - \mu_2
\right) \nonumber \\
\tilde{b} & = & 1 + \frac{1}{2} \left( \frac{{\cal M}}{2\pi} +\mu_1 +\mu_2
\right) \nonumber \\
\tilde{c} & = & 1 +\mu_1 . \label{b33}
\end{eqnarray}
This function is solution of the Fuchsian equation (\ref{a33}) with a
different choice of the index at infinity, namely $|\mu_{\infty}| = 1
+ \frac{{\cal M}}{2\pi}$.

The non perturbative solution can be expressed as:
\begin{eqnarray}
\psi'& = & \Omega_1 (\zeta) ( y_{-} (0) + \overline{f} (0) k_1 y_{+} (0) ) + i
\Omega_2 (\zeta) ( \overline{f} (0) \tilde{y}_{-} (0) +  k_2 \tilde{y}_{+} (0)
) \nonumber \\
\chi'& = & - i \Omega_1 (\zeta) ( f(0) y_{-} (0) + k_1 y_{+} (0) ) 
+ \Omega_2 (\zeta) ( \tilde{y}_{-} (0) + k_2 f(0) \tilde{y}_{+} (0) ) ,
\label{b34} \end{eqnarray}
where the meromorphic functions $\Omega_1 (\zeta)$ and $\Omega_2
(\zeta)$ are defined as 
\begin{eqnarray}
\Omega_1 (\zeta) & = & \frac{\delta_1}{\zeta} +
\frac{\sigma_1}{1-\zeta} \nonumber \\
\Omega_2 (\zeta) & = & \frac{\delta_2}{\zeta} + 
\frac{\sigma_2}{1-\zeta}  ,
\label{b35} \end{eqnarray} 
and $\delta_i, \sigma_i$ are necessary parameters 
to satisfy the translation part of the monodromies.
Instead the coefficients $k_1$ and $k_2$ are determined by
imposing the monodromy conditions for $\psi'$ and $\chi'$ 
around $\zeta=1$. 
Let us introduce new bases around $\zeta = 1$ whose monodromy is
diagonal turning around particle \# 2
\begin{eqnarray}
y_{+} (1) & = & \zeta^{-\frac{\mu_1}{2}}
{(1-\zeta)}^{\frac{\mu_2}{2}}
\widetilde{F} ( a', b', a'+b'-c'+1; 1-\zeta) \nonumber \\
y_{-} (1) & = & \zeta^{-\frac{\mu_1}{2}}
{(1-\zeta)}^{-\frac{\mu_2}{2}}
\widetilde{F} ( a, b, a+b-c+1; 1-\zeta),
\label{b36} \end{eqnarray}
where $(a,b,c)$ are defined as in eq.(\ref{b31}), and
\begin{eqnarray}
\tilde{y}_{+} (1) & = & \zeta^{\frac{\mu_1}{2}}
{(1-\zeta)}^{-\frac{\mu_2}{2}}
\widetilde{F} ( \tilde{a'}, \tilde{b'},
\tilde{a'}+\tilde{b'}-\tilde{c'}+1;1-\zeta) \nonumber \\
\tilde{y}_{-} (1) & = & \zeta^{\frac{\mu_1}{2}}
{(1-\zeta)}^{\frac{\mu_2}{2}}
\widetilde{F} ( \tilde{a}, \tilde{b}, \tilde{a}+\tilde{b}-\tilde{c}+1;
1-\zeta), 
\label{b37} \end{eqnarray}
with  $(\tilde{a},\tilde{b},\tilde{c})$ given by the other choice (\ref{b33}).

It is not difficult to realize that, by using the rules of analytic 
continuation of the hypergeometric functions, $y_{+} (0)$ and $y_{-}
(0)$ can be expressed in the new basis as:
\begin{eqnarray}
y_{+} (0) & = & K \ 
\left( \frac{sin \pi a \ sin \pi b }{sin \pi (c-a) sin \pi (c-b)}
y_{-} (1) - y_{+} (1) \right) \nonumber \\
y_{-} (0) & = & K \left( y_{-} (1) - y_{+} (1) \right) \ \ \ \ \ \ \ \ \  \ 
K =  \frac{\pi}{sin \pi(c-a-b) \Gamma (c-a) \Gamma (c-b) } ,
\label{b38} \end{eqnarray}
where, apart from the common factor $K$, it appears the characteristic factor: 
\beq \frac{sin \pi a \ sin \pi b }{sin \pi (c-a) sin \pi (c-b)} =
\frac{\gamma_{12} -1 }{\gamma_{12}+1} . \label{b39} \eeq
The formula (\ref{b39}) 
is valid for both choices (\ref{b31}) and (\ref{b33}) of the
parameters of the hypergeometric function and therefore for 
both $y_{\pm}$ and $\tilde{y}_{\pm}$,
while $K = K(a,b,c)$ has to be substituted with $\widetilde{K} = K ( \tilde{a},
\tilde{b}, \tilde{c} )$.

By comparing the analytic continuation of the solution near $\zeta=0$
with the expected solution near $\zeta=1$
\begin{eqnarray}
\psi'& = & \widetilde{\Omega}_1 (\zeta) 
( y_{-} (1) + \overline{f} (1) {k'}_1 y_{+} (1) ) + i
\widetilde{\Omega}_2 (\zeta) 
( \overline{f} (1) \tilde{y}_{-} (1) + {k'}_2 \tilde{y}_{+} (1)
) \nonumber \\
\chi'& = & - i \widetilde{\Omega}_1 (\zeta)
( f(1) y_{-} (1) + {k'}_1 y_{+} (1) ) +
\widetilde{\Omega}_2 (\zeta) 
( \tilde{y}_{-} (1) + f(1) {k'}_2 \tilde{y}_{+} (1) ) ,
\label{b40} \end{eqnarray}
we find constraints on $k_i$ and $k_i'$ which, for collinear
velocities, are solved by:
\begin{eqnarray} k_1 \ & = & \ \overline{k}_2 = \ 
\frac{\gamma_{12} \overline{V}_{21}}{\gamma_{12}-1} \nonumber \\
{k'}_1 \ & = & \ \overline{k'}_2 = \ \frac{\gamma_{12} V_{12} }{\gamma_{12}-1}.
\label{b41} \end{eqnarray}
Instead $\Omega_1(\zeta) $ and $\Omega_2(\zeta) $ are computable from
the monodromy conditions for the integrated fields $\psi$ and
$\chi$ as in eq. (\ref{b13}), which allows to determine the
anticommuting parameters in terms of two independent ones.

The singular behaviour of the fields can be easily obtained around
$\zeta =0$
\begin{eqnarray}
\psi & = & \Delta_1 \zeta^{-\frac{\mu_1}{2}} + i \overline{f} (0) 
\Delta_2 \zeta^{\frac{\mu_1}{2}} + O( \zeta^{1-\frac{\mu_1}{2}} )
+ O( \zeta^{1+\frac{\mu_1}{2}} ) \nonumber \\
\chi & = & - i f(0) \Delta_1 \zeta^{-\frac{\mu_1}{2}} + \Delta_2
\zeta^{\frac{\mu_1}{2}}  + O( \zeta^{1-\frac{\mu_1}{2}} )
+ O( \zeta^{1+\frac{\mu_1}{2}} ) \nonumber \\
\Delta_1 & = & - \frac{2}{\mu_1} \left[ \delta_1 
\frac{\Gamma(a) \Gamma(b)}{\Gamma(c)} + i \delta_2 k_2 
\frac{\Gamma(\tilde{a}') \Gamma(\tilde{b}')}{\Gamma(\tilde{c}')}
\right]  \nonumber \\
\Delta_2 & = & \frac{2}{\mu_1} \left[ \delta_2 
\frac{\Gamma(\tilde{a}) \Gamma(\tilde{b})}{\Gamma(\tilde{c})} - i \delta_1 k_1 
\frac{\Gamma(a') \Gamma(b')}{\Gamma(c')}
\right]  \label{b42}
\end{eqnarray}
and around $\zeta=1$
\begin{eqnarray}
\psi & \sim & \tilde{\Delta}_1 {(1-\zeta)}^{-\frac{\mu_2}{2}} 
+ i \overline{f} (1) \tilde{\Delta}_2 {(1-\zeta)}^{\frac{\mu_2}{2}} 
+ O( {(1-\zeta)}^{1-\frac{\mu_2}{2}} ) + O( {(1-\zeta)}^{1+\frac{\mu_2}{2}} )
\nonumber \\
\chi & \sim & - i f(1) \tilde{\Delta}_1 {(1-\zeta)}^{-\frac{\mu_2}{2}}
+ \tilde{\Delta}_2 {(1-\zeta)}^{\frac{\mu_2}{2}} 
+ O( {(1-\zeta)}^{1-\frac{\mu_2}{2}} ) + O( {(1-\zeta)}^{1+\frac{\mu_2}{2}} ) 
\nonumber \\
\tilde{\Delta}_1 & = & \frac{2}{\mu_2} \left[ \sigma_1 H 
\frac{\Gamma(a) \Gamma(b)}{\Gamma(a+b-c+1)} + i \sigma_2 \widetilde{H} {k'}_2 
\frac{\Gamma(\tilde{a}') \Gamma(\tilde{b}')}{\Gamma(\tilde{a}'+
\tilde{b}'-\tilde{c}'+1)}
\right] \nonumber \\
\tilde{\Delta}_2 & = & \frac{2}{\mu_2} \left[ - \sigma_2 \widetilde{H} 
\frac{\Gamma(\tilde{a}) \Gamma(\tilde{b})}{\Gamma(\tilde{a}+\tilde{b}-
\tilde{c}+1)} + i \sigma_1 H {k'}_1 
\frac{\Gamma(a') \Gamma(b')}{\Gamma(a'+b'-c'+1)}
\right] \nonumber \\
H & = & K \left( 1+\overline{f}(0) \frac{\gamma_{12} 
\overline{V}_{12}}{\gamma_{12}+1} \right) \quad \quad \quad \widetilde{H} =
\widetilde{K} \left( 1+ f(0) \frac{\gamma_{12} V_{21}}{\gamma_{12}+1} \right). 
\label{b43} \end{eqnarray}
The logarithmic behaviour of $T$ and $Z$ variables is constrained by
the property that $Z-V_i T$ are free of singularities, and therefore
it is enough to compute the one for $T$ around each particle
\begin{eqnarray} \Delta_S T & {\buildrel 1 \over \rightarrow} &
\Delta_S T  + 4\pi \mu_1 
Re (\Delta_1 \Delta_2) ( 1 + f(0) \overline{f} (0) )
 \nonumber \\ \Delta_S T & {\buildrel 2 \over \rightarrow} & 
\Delta_S T + 4 \pi \mu_2 
Re (\widetilde{\Delta}_1 \widetilde{\Delta}_2) 
( 1 + f(1) \overline{f} (1)) . \label{b44} \end{eqnarray}
By solving the system of conditions we find that $\Omega_1 (\zeta)$
and $\Omega_2 (\zeta)$ are of order $O(V^2)$. Then the static limit is
recovered because $y_{-} (0)$ and $\tilde{y}_{-} (0)$ scale as 
$O ( 1 /V^2 )$ due to the vanishing of the $a$ parameter in the 
hypergeometric function $\widetilde{F} (a, b, c, \zeta)$.

By putting together the solution for gravity, described in Sec. $2$,
and the exact solution for the gravitino fields, given in eq. (\ref{b34}),
we have complete control of the fields. Moreover the fermionic
trajectory is already determined in terms of the
bosonic one by eq. (\ref{b26}):
\beq \theta(0) = - \frac{\Delta_1}{C} \xi^{\frac{{\cal M}}{4\pi}} \quad
\quad \quad \theta(1) = - \frac{\widetilde{\Delta}_1}{C} 
\xi^{\frac{{\cal M}}{4\pi}} . \label{b45} \eeq
Let us note that the difference of fermionic coordinates in this case
may depend on the residual arbitrariness of the anticommuting parameters
which persists after solving the monodromy conditions. However, it is 
possible to add homogeneous solutions of the monodromies, depending on
arbitrary parameters $\alpha_i = \alpha_i(\mu_j, {\cal M})$, $\beta_i
= \beta_i(\mu_j,{\cal M}) $, to the gravitino fields
\begin{eqnarray}
\tilde{\psi}
 & = & \psi + \Delta \psi \qquad \qquad \qquad \Delta \psi \rightarrow
\overline{a}_i \Delta \psi + i \overline{b}_i \Delta \chi \nonumber \\
\Delta \psi & = & ( \alpha_1 \delta_1 + \beta_1 \sigma_1 )
[ y_{-} (0) + \overline{f} (0) k_1 y_{+} (0) ] + i 
( \alpha_2 \delta_2 + \beta_2 \sigma_2 )
[ \overline{f} (0) \tilde{y}_{-} (0) + k_2 \tilde{y}_{+} (0) ]
\nonumber \\
\tilde{\chi}
 & = & \chi + \Delta \chi \qquad \qquad \qquad \Delta \chi \rightarrow
a_i \Delta \chi - i b_i \Delta \psi \nonumber \\
\Delta \chi & = & - i ( \alpha_1 \delta_1 + \beta_1 \sigma_1 )
[ f(0) y_{-} (0) + k_1 y_{+} (0) ] + 
( \alpha_2 \delta_2 + \beta_2 \sigma_2 )
[ \tilde{y}_{-} (0) + k_2 f(0) \tilde{y}_{+} (0) ] \nonumber \\
& & \label{b46} \end{eqnarray} 
which do not change the good properties of the solution but instead 
are able to connect $\theta(1) -\theta(0)$ only to a linear combination of
the constants of motion. 

It remains to be investigated the equation for $\xi (t)$ which solves
the whole motion of superparticles. At level of the bosonic trajectory
\beq B_2 - B_1 = Z_2 - Z_1 - V_2 T_2 + V_1 T_1  , \label{b47} \eeq
we expect to find finite contributions derived from the $\xi_i^F$
fermionic trajectory. Let us start to analyze the $T$-mapping. 
In this case the finite term due, for example,  to $\xi^F_1$ is given, in the
normalization of eq. (\ref{b21}) by the term
\beq \theta  \left( i \chi + \psi f \right) \sqrt{\frac{N}{f'}}|_0
= i ( 1 + \overline{f} (0) f(0) ) \Delta_2 \Delta_1 + (h.c.),
\label{b48}  \eeq
which is obtained by a compensation of $\xi^{\pm \mu_i}$ exponents.

In the $Z$-mapping there is also a finite contribution due  to
$\xi^F_1$ which might enter into the geodesic equations (\ref{b47}) 

\beq 2 \theta \psi \sqrt{\frac{N}{f'}} |_0 - 
2i \overline{\chi} \sqrt{ \frac{\overline{N}}{\overline{f}'} }
\overline{f} \overline{\theta} |_0 = 2 \overline{f} (0) \ ( i \Delta_2
\Delta_1 + ( h.c. ) ) . \label{b49} \eeq

It is interesting to observe that, by taking the combination
$Z(\xi_i) - V_i T(\xi_i)$, these finite 
contributions cancel exactly due again to eq. (\ref{b23}) and 
from now on we can limit our discussion to the $\theta = 0$ part of
eq. (\ref{b47}).

However there is another obstacle to consider, namely the logarithmic
divergences which make the geodetic equations (\ref{b47}) undetermined.
To define them properly avoiding the logarithmic ambiguities,
it is enough to compare the two superparticles from a fixed reference point
$C$. For simplicity we take $C$ in a fixed position in the
$\zeta$-plane, since then it is easy to compare the $\xi$-dependence
of the various terms which contribute to the geodesics. 

Since the combinations $Z(\xi_i) - V_i T(\xi_i)$ are well defined, it is
convenient to compute the following building block of the geodesics equations
( without the $\theta$ contributions ):
\begin{eqnarray}
( Z(\xi_i) & - & Z(C) - V_i T(\xi_i) + V_i T(C) ) |_{\theta=0} = 
\int^{\xi_i}_C \ \left( \frac{N}{f'} - 2 \psi {\psi'}
\right) dz + \nonumber \\
& + & \int^{\overline{\xi}_i}_{\overline{C}} \ \left( 
\frac{\overline{N}}{\overline{f}'} \overline{f}
- 2 \overline{\chi} {\overline{\chi}'}
\right) d{\overline z} - V_i \left[ 
\int^{\xi_i}_C  \ dz \ \left( \frac{N f}{f'} -
i ( \chi {\psi'} + \psi {\chi'} \right) + (h.c.) 
\right] \ \ \ \ \ \ \ \nonumber \\
& & (i=1,2).
\label{b50} \end{eqnarray}

Since $\psi(x)$ depends only on $\zeta$, all the fermionic integrals
scale as a constant ($\xi^0$), while the bosonic integral scales as
$\xi^{1-\frac{{\cal M}}{2\pi}}$. Therefore, for large times, the pure bosonic
contribution dominates ( the gravitino has shorter range than the
graviton ).  By taking the difference of the building blocks of eq. 
(\ref{b50})
we obtain a regularized version of eq. (\ref{b47})
\begin{eqnarray}
Z(\xi_2) & - & V_2 T(\xi_2) - Z(\xi_1) + V_1 T(\xi_1) = B(\xi_2) -
B(\xi_1) = \nonumber \\
& = & (V_1 - V_2) T(C) + \int^{\xi_2}_{\xi_1} \left( \frac{N}{f'} - 
2 \psi {\psi'} \right) dz +
\int^{\overline{\xi}_2}_{\overline{\xi}_1} \left( \frac{\overline{N}}{
\overline{f}'} \overline{f} - 2 \overline{\chi} 
{\overline\chi'} \right) d\overline{z}  \nonumber \\
& - & V_2 \left[ \int^{\xi_2}_C \ dz \ \left( \frac{N f}{f'} -
i ( \chi {\psi'} + \psi {\chi'} \right) + (h.c.) \right] + \nonumber \\
& + &  V_1 \left[ \int^{\xi_1}_C \ dz \ \left( \frac{N f}{f'} -
i ( \chi {\psi'} + \psi {\chi'} \right) + (h.c.) \right].
\label{b51} \end{eqnarray}
We have eliminated $Z(C)$, and we are left with
$T(C)$ which is the freedom of choosing an universal time, up to 
time reparametrizations. For large time $T(C) \sim t$, while it is clear
that the dominant term, in the remaining part, is given by the pure
bosonic contribution. Therefore it is possible to define a scattering
angle at the classical level and it coincides with the result of the
gravity case \cite{ciaf}. This was also expected at the level of the monodromy 
matrices $L_i$, since the scattering angle is strictly related to the 
invariant mass of the system which is computable from the Lorentz part of the 
composite monodromy $L_{12} = L_1 L_2$, while the supersymmetric cut gives
contribution only to the translation part. 

The main effect of the anticommuting variables is to
introduce a certain level of unpredictability of the trajectory, but
not of the scattering angle. To introduce probabilities for the 
scattering angle we need quantum mechanics \cite{quan}. 

\subsection{Extensions}

A natural extension would be to apply this method to the study of the
$N$ body problem. We have already investigated it in the case of point
particles in $(2+1)$ gravity. In that case the solution is found by
requiring that the function $f(z,t)$ respects the projective transformations
(\ref{a32}) around each particle. Since the monodromy matrices are constants of
motion, we have to deal with an isomonodromy problem \cite{isom}, 
which is solved
at level of $f$ with the introduction of apparent singularities in a
Fuchsian differential equation analogous to 
(\ref{a33}). The isomonodromy then leads to constraint
the motion of apparent singularities in terms of the physical ones.
Therefore since the monodromy problem for $\psi$ and $\chi$ is very
similar to the one for $f(z,t)$ one expects that apparent
singularities must be present in their solution too.
From the two-body problem we can deduce that their general solution
can be represented as a sum of two distinct solutions of a Fuchsian
differential equation, having $(N+1)$ physical poles and $N-2$
apparent singularities. The two solutions are different 
because all masses enter with opposite sign, and in particular
this is also true for the total mass of the system
${\cal M}$ related to the loop invariant $L_1 L_2 ... L_N$. Therefore in
one case the index at infinity has to be defined as $\mu_\infty =
\frac{\cal M}{2\pi}-1$, in the other case as $ | \mu_\infty | =
\frac{\cal M}{2\pi}+1$.

We will not proceed further in this direction. In any case, it seems
to us obvious that a solution to the monodromy problem (\ref{a36})
must exists for $N$ superparticles. Instead we would like
to investigate the static case for three bodies ( $z= \xi_i$,
$i=1,2,3$ ) , in order to clarify the procedure for obtaining the
static solution, which gives an idea of the parameterization of the 
interacting solution. In the static limit the $\psi$ function can be 
expressed, by choosing an adimensional variable 
$\zeta = ( z-\xi_1)/(\xi_2-\xi_1)$ and $\xi = \xi_{13} / \xi_{12}$, as follows

\begin{eqnarray} {\psi'} (\zeta) & = & \left( \frac{\delta_1}{\zeta} + 
\frac{\sigma_1}{\zeta-1} +
\frac{\rho_1}{\zeta-\xi} \right)
\ \zeta^{-\frac{\mu_1}{2}} {(\zeta-1)}^{-\frac{\mu_2}{2}}
 {(\zeta-\xi)}^{-\frac{\mu_3}{2}} . \label{b52} 
\end{eqnarray}
Around the superparticles the singular behaviour of the
$\Theta$-mapping, at the level of $\theta$-terms, constraints the
fermionic geodetics to be:
\beq \theta (0)  = - \ \frac{2}{\mu_1} \ \delta_1 \  
\xi_{12}^{\frac{\mu_1+\mu_2+\mu_3}{2}} \ \ \ 
\theta (1)  = - \ \frac{2}{\mu_2} \ \sigma_1 \  
\xi_{12}^{\frac{\mu_1+\mu_2+\mu_3}{2}} \ \ \ 
\theta (\xi)  = - \ \frac{2}{\mu_3} \ \rho_1 \
\xi_{12}^{\frac{\mu_1+\mu_2+\mu_3}{2}} . \ \ \ \label{b53} 
\eeq
Analogously, $\chi'(\zeta)$ can be defined as:
\begin{eqnarray} {\chi'} (\zeta) & = & \left( \frac{\delta_2}{\zeta} + 
\frac{\sigma_2}{\zeta-1} + \frac{\rho_2}{\zeta-\xi} \right)
\ \zeta^{\frac{\mu_1}{2}} {(\zeta-1)}^{\frac{\mu_2}{2}}
{(\zeta-\xi)}^{\frac{\mu_3}{2}}. \label{b54} 
\end{eqnarray}
Imposing the monodromy conditions 
\begin{eqnarray} \psi (\xi) - \psi(0) & = & \int^{\xi}_0 d\zeta \
\psi' \quad \quad \quad 
\chi (\xi) - \chi(0) = \int^{\xi}_0 d\zeta \chi' \nonumber \\ 
\psi (1) - \psi(0) & = & \int^1_0 d\zeta \psi' \quad \quad \quad 
\chi (1) - \chi(0) = \int^1_0 d\zeta \chi' .  \label{b55}
\end{eqnarray}
allows to compute four anticommuting unknowns in terms of two
independent ones. Moreover the differences of coordinates
$\theta(\xi_i) - \theta(\xi_j)$ are always dependent only on the
constant of motions, as in the two-body static case.
In general, for a static $N$ body problem, formulas similar 
to eqs. (\ref{b52}) and (\ref{b54}) hold carrying $N$ anticommuting parameters.

Finally, we expect that solving the $N$-body problem in supergravity
is no more difficult than in gravity, where, although the non
perturbative $N$-body solution cannot be given in terms of known
tabulated functions, it is possible
to construct solutions to the monodromy matrices, at least
perturbatively in $V_i$ \cite{ciaf}.

Another straightforward extension of this formalism is to understand
extended supergravity. For example, $N=2$ supersymmetry has two
$\Theta_i$, superspace variables, and the corresponding supermetric is 
\begin{eqnarray}
ds^2 & = & {[ dT + \frac{i}{2} ( \Theta_i d \overline{\Theta}_i +
\overline{\Theta}_i d \Theta_i ) ]}^2  - \nonumber \\
& - & ( dZ  + \Theta_i d \Theta_i ) ( d
\overline{Z} + d\overline{\Theta}_i \overline{\Theta}_i ) - 
( d \Theta_i d \overline{\Theta}_i ) .
\label{b56} \end{eqnarray}

Besides the supersymmetric translation cuts for $\Theta_i$, it is also
possible to have cuts related to the
mixing of the two $\Theta_i$ which however must leave the bilinears
$\Theta_i d \Theta_i$, and $\Theta_i d \overline{\Theta}_i +
\overline{\Theta}_i d \Theta_i$ invariant. In any case, the bosonic variables
($T,Z$) feel only the supersymmetric cuts $\Theta_i \rightarrow
\Theta_i + \epsilon_i$.
Straightforward generalizations of our parameterization (\ref{a37}) allow to
solve any kind of supersymmetric $N$-body problem always in terms of 
analytic functions.

\section{Conclusion}

In this paper we have been able to solve explicitly the
non-perturbative interaction between $N$ superparticles and supergravity.
We have analyzed in detail the two-superbody system in our
coordinates with instantaneous propagation, and sketched
the classical $N$-body problem. Our method consists of the following steps.
Firstly we have introduced a new parameterization for the $X$-supermapping,
which extends the well known superanalytic mapping, and gives the
general solution to the equations of motion of pure supergravity in
terms of arbitrary analytic functions. 

Introducing specific polydromies in the mapping we can couple in a
consistent way $N$ superparticles to $(2+1)$ supergravity. The
bootstrap between motion of superparticles and field equations is
achieved by imposing constraints which are equivalent to integrate the
supergeodetic equations. In addition to evaluate the bosonic distance
of the superparticles, a criterion which was known from the gravity
problem, we have found, by examining in detail the explicit solution
of the supermapping, a second fermionic constraint. It represents the
odd part of the supergeodesic equations and gives a precise meaning to a non
perturbative divergence at the particle site, which was already
present in the case of spinning particles coupled to gravity.

The requirement of cancellation of the fermionic residue at this
divergence is found to be consistent at all orders, by using only
general properties of the monodromy matrices. This second constraint
is an interesting outcome of this work, which
was difficult to carry out because of the lack of explicit examples
available with anticommuting variables.

Finally, we have learned that divergences cannot be considered
negligible, as we usually do in field theory, but have a important
role to build a more complete theory.  

{\bf Acknowledgements }

I would like to thank Dr. A. Bellini and Prof. M. Ciafaloni for useful
discussions.

\appendix
\section{Appendix - Two-body solution in ($2+1$)-gravity}

In this Appendix we review the solution of $(2+1)$-gravity. To build a
representation of the matching conditions around the particles,
i.e. to impose that for $(z - \xi_i) \rightarrow e^{2\pi i} \ (z - \xi_i)$

\beq dX^a \rightarrow  \ {(L_i)}^a_b dX^b ,  \label{c1} \eeq
we introduced the analytic function $f(z,t)$, which has branch points at
$z \ = \ \xi_i(t)$ such that, when $z$ turns around $\xi_i$, $f$
transforms as a projective representation of the monodromies 

\beq f(z,t) \rightarrow \frac{a_i f(z,t) + b_i}{{\bar b}_i f(z,t) + {\bar a}_i}
\ , \ \ \ ( i = 1, ... , N ), \label{c2} \eeq

with 
$$ a_i \ = \ \cos \frac{m_i}{2} + i \gamma_i \sin \frac{m_i}{2} \ , \ \ \ \ \ 
b_i \ = \ - i \gamma_i {\bar V}_i \sin \frac{m_i}{2} , $$
\beq \gamma_i \equiv {( 1 - {|V_i|}^2 )}^{-1/2} , \ \ \ \ \ 
V_i \ = \ ( P^x_i + i P_i^y )/E_i  \ . \label{c3} \eeq

The specific form of $f(\zeta,t)$ for the two-body problem is given by

\beq f(\zeta,t)  =  \frac{f(0) + f_{(1)}}{1+\overline{f}(0) f_{(1)} } .
\label{c4} \eeq
The value $f(0)$ is  determined by the $L_1$ monodromy matrix
\beq f(0) = \frac{\gamma_1 \overline{V}_1}{1+\gamma_1},  \label{c5} \eeq
and it corresponds to the fixed point of the projective
transformation (\ref{c2}). $f_{(1)}$ would be the solution for the
monodromies (\ref{c2}) if we decide to see the particle scattering in the
particle \#1 rest frame:
\beq f_{(1)} (\zeta) \ = \ \frac{\gamma_{12} {\bar V}_{21}}{\gamma_{12} - 1}
\frac{\zeta^{\mu_1} \ {\widetilde F} (a',b',c';\zeta)}{{\widetilde F} 
(a,b,c;\zeta)} . \label{c6} \eeq
In fact, the monodromy around particle \#1 becomes a pure rotation .
To derive this formula we have assumed for simplicity that
the velocities are collinear, since then the relativistic difference
of velocities simplifies to
\beq V_{12}= \frac{V_1 - V_2}{1-\overline{V}_1 V_2} , \label{c7} \eeq  and  
$\gamma_{12} \equiv (P_1 P_2)/ m_1 m_2 $ is the relative
$\gamma$-factor. The factor in front to the ratio of the
hypergeometric functions is chosen to satisfy the monodromy conditions
around particle \#2, and in particular to obtain the value 
\beq f(\zeta = 1) = \frac{\gamma_2 \overline{V}_2}{1+\gamma_2} .
\label{c10} \eeq

In the formula (\ref{c6}) we have introduced a modified hypergeometric
function solution of the hypergeometric equation (\ref{a33})
\beq \widetilde{F} (a,b,c;z) \equiv \frac{\Gamma(a) \Gamma(b)}{\Gamma(c)} 
F(a,b,c;z)  . \label{c8} \eeq

The choice of the parameteres of the hypergeometric function is defined
as follows:
\begin{eqnarray} a(\mu_1) & = & \frac{1}{2} 
( 1 + \mu_\infty - \mu_1 - \mu_2 ), 
\ \ b(\mu_1) \ = \ \frac{1}{2} 
( 1 - \mu_\infty -\mu_1 - \mu_2 ) , \ c(\mu_1) \ = \ 
1 - \mu_1  \nonumber \\
 a' & = &  a(-\mu_1) , \ b' \ = \ b(-\mu_1), \ c' \ = \ c(-\mu_1) . 
\label{c9} \end{eqnarray}
The values of the indices $\mu_i$ are defined in eq. (\ref{a34}).

The two hypergeometric function have a natural definition of their
cuts which entails the structure of cuts for $f(z,t)$, which split
the $z$-plane in two half-planes, apart from a segment of continuity on
the real axis between $0$ and $1$. In the mass range 
\beq 0 \leq m_1, m_2 , {\cal M} \le 2\pi \label{c11}  , \eeq
the upper half $z$-plane is mapped on a triangle whose edges are
circular arcs, and whose internal angles are $\frac{m_1}{2}$,
$\frac{m_2}{2}$ and $\pi - \frac{{\cal M}}{2}$, for $m_1 + m_2 \le
{\cal M} \le 2\pi$. The lower half plane is obtained by Schwarz's
reflection of this triangle. The whole $z$-plane is thus mapped to a region
inside the $f$ unit disk, $|f(z,t)| \le 1$, and the same inequality is
satisfied by $f(z,t)$ on any other Riemann sheet.

In particular $f(z,t)$ at $\infty$ has two values, one for each half plane,
$f_{+} (\infty) (f_{-} (\infty) )$, which are fixed points of the 
composite loop operators $L_{21} ( L_{12} )$. Until the constant
velocities do not saturate the Carlip bound 
$cos \frac{{\cal M}}{2} < -1$, the value of $f_{\pm} (\infty)$ is
always contained into the $f$-unit disk and the determinant of the
metric is always non vanishing. This
property extends to the supergravity case, since the
anticommuting parameters do not play a significant role in inverting
the supermetric, until there is a non vanishing pure bosonic part.

%****************************************************************************

%****************************************************************************

\end{document}